%% file: mainrevtexv2.tex
\documentclass[aps,prd,twocolumn,showpacs,preprintnumbers,amsmath,amssymb]{revtex4-1}
\usepackage{amsmath}
\usepackage{graphicx}
\usepackage{float}
\usepackage{caption}
\usepackage{fancyref}
\usepackage{subcaption}
\usepackage{tikz}
\usepackage{subfiles}
\usepackage{float}
\usepackage{siunitx}
\usepackage{wrapfig}
\usepackage{xargs}      
\usepackage[colorlinks]{hyperref}
\definecolor{winered}{rgb}{0.8,0,0}
\definecolor{darkb}{rgb}{0,0,0.8}
\hypersetup{
    colorlinks=true,
    citecolor=blue,
    linkcolor=winered,
    filecolor=red,      
    urlcolor=darkb,
}
\newcommand{\ket}[1]{|#1 \rangle}
\newcommand{\bra}[1]{\langle #1 |}
\newcommand{\onlyinsubfile}[1]{#1}
\newcommand{\notinsubfile}[1]{}
 
\usetikzlibrary{backgrounds,fit,decorations.pathreplacing}

\begin{document}
\title{Quantum simulation of scattering in the quantum Ising model 
}
\author{Erik Gustafson$^1$}
\author{Y. Meurice$^1$}
\author{Judah Unmuth-Yockey$^2$}
\affiliation{$^1$ Department of Physics and Astronomy, The University of Iowa, Iowa City, IA 52242, USA }
\affiliation{$^2$Department of Physics, Syracuse University, Syracuse, NY 13244 USA}
\date{\today}
\renewcommand{\onlyinsubfile}[1]{}
\renewcommand{\notinsubfile}[1]{#1}
\begin{abstract}
We discuss real time evolution for the quantum Ising model in one spatial dimension with $N_s$ sites. In the limit where the nearest neighbor interactions $J$ in the spatial directions are small, 
there is a simple physical picture where qubit states can be interpreted as approximate particle occupations.  Using exact diagonalization, for initial states with one or two particles, we show that for small $J$, discrete Bessel functions provide very accurate expressions for the evolution of the occupancies corresponding to initial states with one and two particles. Boundary conditions play an important role when the evolution time is long enough. We discuss a Trotter procedure to implement the evolution on existing quantum computers and discuss the error associated with the Trotter step size. We discuss the effects of gate and measurement errors on the evolution 
of one- and two-particle states using 4 and 8 qubits circuits approximately corresponding to existing or near term quantum computers. 
 
\end{abstract}
\maketitle

\section{Introduction}
There has been a fast growing interest for quantum computation in the context of high energy and nuclear physics \cite{Jordan:2011ne,
Martinez:2016aa,Moosavian:2017tkv,Klco:2018kyo,
Lamm:2018siq,Dumitrescu:2018njn,Macridin:2018oli,Raychowdhury:2018osk,Stryker:2018efp,
Yeter-Aydeniz:2018mix,Hackett:2018cel,Klco:2018zqz,Roggero:2018hrn,Muschik:2016tws,Kokail:2018eiw,Lu:2018pjk,Somma2016QuantumSO}.
One important motivation is to calculate real-time evolution of states in large Hilbert spaces which cannot be handled with standard sampling methods. The long term goals include jet physics and early
cosmology. However, in the near term it is important to demonstrate that it is possible to make progress towards these major goals
using quantum computers or quantum simulation experiments with a limited number of qubits \footnote{We currently are discussing the regime of 4 to 8 qubits}.

Numerical lattice gauge theory started in the late 70's by studying $Z_2$ (Ising) gauge theories on $3^4$ lattices and has
steadily developed as a reliable tool that today allows different collaborations to compare numerical estimates for hadronic processes with
errors of a few percent.  
It thus seems natural to start the study of real time evolution using the quantum
Ising model in 1+1 dimension \cite{Lamm:2018siq} or the Schwinger model \cite{Martinez:2016aa,Klco:2018kyo}.

In the following we propose to consider simple cases of time evolution for the quantum Ising model with a number of sites of the same order
as the number of qubits in devices existing  or expected to exist in the near future. The model has a second order phase transition
which allows the use of finite size scaling (FSS) to extract interesting information using systems with a small number of lattice sites. This strategy is explained in Ref.~\cite{PhysRevLett.121.223201}. The main goal is to provide reliable benchmark calculations in situations that will allow comparison among different platforms.

The paper is organized as follows. In Sec. \ref{sec:ham}, we present the model and  in Sec. \ref{sec:dpt} we present some perturbative results. In Sec. \ref{sec:realtime} the Suzuki-Trotter formulation of the time evolution operator for the quantum Ising model is derived. In Sec. \ref{sec:results} we discuss how artificial noise is introduced into our simulations as well as the results of our simulations for both free propagation and ``scattering" of particles. We follow the methodology inspired by the one laid out in Ref. \cite{Jordan:2011ne}. We first prepare highly localized ``wavepackets" which can be considered a two-particle state. Then we let the packets spread and ``interact" using the corresponding Hamiltonian. 
\section{The Quantum Ising Model}
\label{sec:ham}
\subsection{Hamiltonian and boundary conditions}
The one-dimensional quantum Ising model is the standard example of a quantum field theory with continuous time that is obtained from a classical lattice model with one extra dimension corresponding to the Euclidean time \cite{PhysRevD.17.2637,RevModPhys.51.659}. In this example, the classical model is the usual two-dimensional Ising model solved by Onsager  \cite{PhysRev.65.117} and Kaufman \cite{PhysRev.76.1232}. The Hilbert space of the quantum model is a tensor product of qubits and the connection to quantum computing is immediate (see Eq. (\ref{eq:sample}) for an illustration)

The connection to the classical model makes the choice of a basis where the nearest neighbor interactions in the spatial direction are diagonal very natural. 
We call this choice the ``spin basis". For reasons that will become clear soon, we use a representation where the other term, often referred to as the 
transverse magnetic field term, is diagonal.  We call this choice the ``particle basis." The two representations are connected by a Hadamard unitary transformation.

In the particle basis, the nearest neighbor interactions (particle hopping) use adjacent pairs of $\hat{\sigma}^x$, governed by $J$, while the transverse magnetic field interactions (on-site coupling) governed by $h_T$ use $\hat{\sigma}^z$, where $\hat{\sigma}^z$ and $\hat{\sigma}^x$ are the traditional Pauli matrices. In the particle basis, we define  the ``particle number" at each $l$ site as
\begin{equation}
    \hat{n}_l = (1 - \hat{\sigma}^z_l)/2.
\end{equation}
We will use these quantum numbers to specify the Hilbert space which is a direct product of two-dimensional qubit spaces at each of the $N_s$ spatial sites. 
Just to give an example for $N_s=4$, the action of a sample operator on a sample state can be illustrated as 
\begin{equation}
\hat{\sigma}^x_3\ket{1011}=\ket{1001}.
\label{eq:sample}
\end{equation}
We will see that the Hamiltonian only connects states for which the total particle number ($\hat{n}=\sum_i  \hat{n} _i$) is the same modulo 2. 

We define the Hamiltonian corresponding to open boundary conditions, hereinafter referred to as (OBC), as, 
\begin{equation}
\label{eq:hamiltonianobc}
H_{obc} = - J \sum_{i=1}^{N_s-1} \hat{\sigma}^x_i \hat{\sigma}^x_{i+1} - h_T \sum_{i=1}^{N_s} \hat{\sigma}^z_i.
\end{equation}
The Hamiltonian corresponding to periodic boundary conditions (PBC) is defined as,
\begin{equation}
\label{eq:hamiltonianpbc}
H_{pbc} = H_{obc} - J\hat{\sigma}^x_{1} \hat{\sigma}^x_{N_s}, 
\end{equation}
while it is also interesting to consider a Hamiltonian with antiperiodic boundary conditions (ABC), \begin{equation}
\label{eq:hamiltonianabc}
H_{abc} = H_{obc} +J\hat{\sigma}^x_{1} \hat{\sigma}^x_{N_s}.
\end{equation}
 For any of these Hamiltonians we define an operator to carry out the exact time evolution in units where $\hbar$ =1 as,
\begin{equation}
\label{eq:evolutionoperator}
U(t) = e^{-i t H}.
\end{equation}

\subsection{Symmetries}
The model has a $Z_2$ global symmetry corresponding to flipping all the spins in the spin basis. In the particle basis, this corresponds to 
multiplying the states by $\hat{\sigma}^z$ at each site. This defines a unitary transformation that flips the sign of the operator $\hat{\sigma}^x$ at each site. 
As such operators come in nearest neighbor pairs, the transformation leaves the Hamiltonian invariant regardless of boundary conditions. This is equivalent to saying that the particle number $\hat{n}$ defined above is conserved modulo 2.

\begin{figure}
    \centering
    \includegraphics[width=0.45\textwidth]{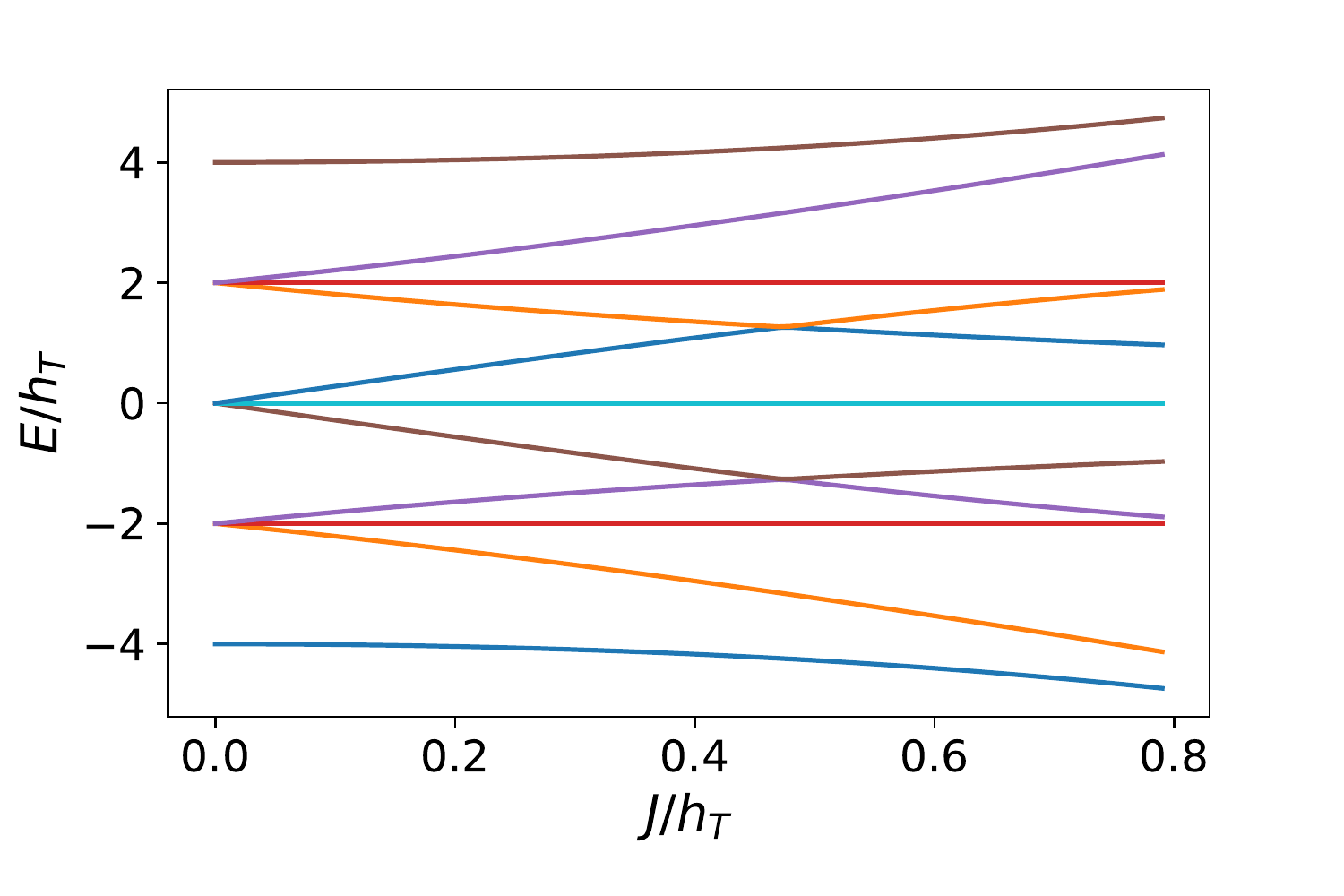}
   \caption{Spectrum for $N_s=4$ in units of $h_T$. Here $h_T$ is the transverse magnetic field and $J$ is the strength of the nearest-neighbor interaction. As $J/h_T$ increases the degenerate energy levels split.}
    \label{fig:spectrum4}
\end{figure}

When $N_s$ is even, it is also possible to invent a two-step transformation which changes the sign of the entire Hamiltonian.  
Similar equivalences appear for classical gauge theories \cite{Li:2004bw}. 
We first apply a $\hat{\sigma}^x$ transformation at each site. This changes the sign of the on-site term and leaves the hopping term unchanged. In a second step, we  apply a   $\hat{\sigma}^z$ {\it on every other site}. The full transformation flips the sign of both terms of the Hamiltonian. Consequently, all the states appear in pairs with opposite signs. This property appears clearly in Fig. \ref{fig:spectrum4} for $N_s=4$. In this case, the 16 states
split into approximately degenerate groups of  1 (0 particle), 4 (1 particle), 6 (2 particles), 4 (3 particles or 1 hole), and 1 (4 particles) when $J<<h_T$. 
In the next section, we discuss the splitting using degenerate perturbation theory. Perturbation theory allows for very accurate calculations of real-time evolution when $J$ is sufficiently small.

\section{Approximate evolution for $J<<h_T$}
\label{sec:dpt}

\subsection{Approximate particle description}
In the limit where  $J = 0$, we obtain a very simple picture for the quantum Ising model. The energy is then the sum of the on-site energies. We have a unique ground state where all sites have an  energy $-h_T$ and so $E^{(0)} = - N_s h_T$. We now have degenerate ``one-particle" states where one on-site state with energy $+h_T$ can be placed at $N_s$ locations. If the $h_T$ energy is located at the site $j$, we call this state $|j \rangle$. These states have an energy $-(N_s - 2)h_T$. Similarly we have ${N_s!}/({n! (N_s-n)!})$ totally antisymmetrized states with $n$ ``particles" and an energy $(-N_s + 2n)h_T$. 
The effect of the nearest neighbor interactions can be included perturbatively \cite{RevModPhys.51.659}. 
The model can also be solved exactly by performing a Wigner-Jordan transformation \cite{PhysRev.76.1232}. However, at finite volume, boundary conditions should be 
treated carefully. To be more explicit, the term $a_{Ns}^\dagger a_1$ needs to be supplemented with a product of $\hat{\sigma}^z_l$ in order to reproduce the original spin Hamiltonian which requires a separate discussion for the even and odd sectors \cite{PhysRev.76.1232}.

\subsection{One particle}

\begin{figure*}
	\begin{subfigure}[H]{0.45\textwidth}
    		\includegraphics[width=\textwidth]{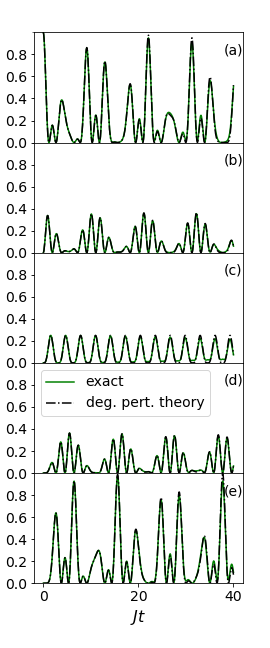}
   	 \end{subfigure}
    	\begin{subfigure}[H]{0.45\textwidth}
    		\includegraphics[width=\textwidth]{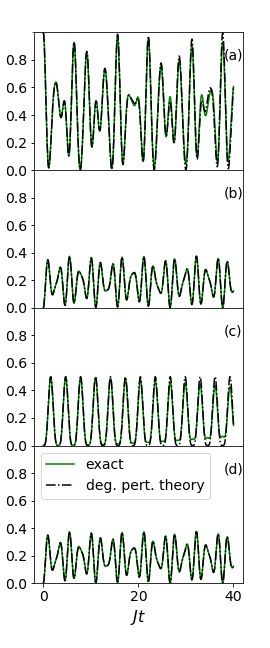}
	\end{subfigure}
		\caption{Comparison of exact diagonalization and perturbation theory. Left: one particle with PBC, where $J = 0.02$, $h_T = 1.0$ and $N_s = 8$. (a) site 1, (b) sites 2 and 8, (c) sites 3 and 7, (d) sites 4 and 6, (e) site 5. Right: two particles with ABC, where $J=0.02$, $h_{T}=1.0$ and $N_s = 8$, (a) sites 1 and 5, (b) sites 2 and 6, (c) sites 3 and 7, and (d) sites 4 and 8. Small discrepancies are most visible in the bottom of (c) on the graphs on the right side.}
		\label{fig:perturbationcompare}
\end{figure*}
At order $J$ in the one-particle sector we have a particle hopping that stays in the one-particle sector. It is worth noting that the particle conservation picture makes this model in the small $J$ limit equivalent to the XY model which has been studied thoroughly. In particular the eigenstates and energies of this model are discussed in Refs. \cite{PhysRev.127.1508,doi:10.1143/ptp/5.1.1} and the zero field case is examined in Refs. \cite{Lieb:1961aa,PERK1977265,2017arXiv171003384p}, time dependent $zz$ spin correlations were studied in Refs. \cite{NIEMEIJER1967377,KATSURA197067} and $xx$ spin correlations were studied in Ref. \cite{Perk2009}.  If periodic boundary conditions are imposed, as in  Eq. \eqref{eq:hamiltonianpbc}, Fourier modes diagonalize the perturbation. This lifts the degeneracy by a term proportional to $2J \text{cos}(2\pi m/N_s)$. The perturbation also contains operators that connect to the 3-particle states; this leads to energy shifts $\mathcal{O}(J^2/h_T)$. If we neglect these second order effects, we have a simple approximate quantum mechanical behavior.

We can then prepare the system in an initial state $|\psi\rangle$ and calculate $\langle \psi(t)| \hat{n}_l |\psi(t) \rangle$, where the calculations in the quantum mechanical approximation are relatively easy. For instance, for $|\psi(0)\rangle = |j\rangle$, we obtain
\begin{equation}
    \label{eq:firstorderocc}
    \langle \psi_j(t)|\hat{n}_l |\psi_j(t)\rangle \simeq |J^{(N_s)}_{l-j}(2Jt)|^2,
\end{equation}
where the ``discrete" Bessel functions are defined as,
\begin{equation}
J_n^{(N_s)}(x) = \frac{(-i)^n}{N_s}\sum_{m=0}^{N_s - 1} e^{i((\frac{2\pi m n}{N_s} + x\text{cos}(\frac{2\pi m}{N_s})))}
\end{equation}
which corresponds to the usual definition in the limit of large $N_s$. In fact these ``Bessel" functions appear in the XY model for the $zz$ correlations as shown in Ref. \cite{KATSURA197067}. The approximation is accurate when $t$ is less than $\mathcal{O}( h_T/(J^2))$ (see Fig. \ref{fig:perturbationcompare}). The implication of this is that pair creation in this model is driven by the hopping parameter $J$, rather than the size of the model. The peaks in the occupation values shown in Fig. \ref{fig:perturbationcompare} suggest that for $J = 0.02$ perturbation theory would be accurate for a system up to 32 sites because approximately 4 resurgences happen before noticeable discrepancies begin to appear. 

\subsection{Two particles}
\begin{figure}[H]
    \centering
    \includegraphics[width=0.45\textwidth]{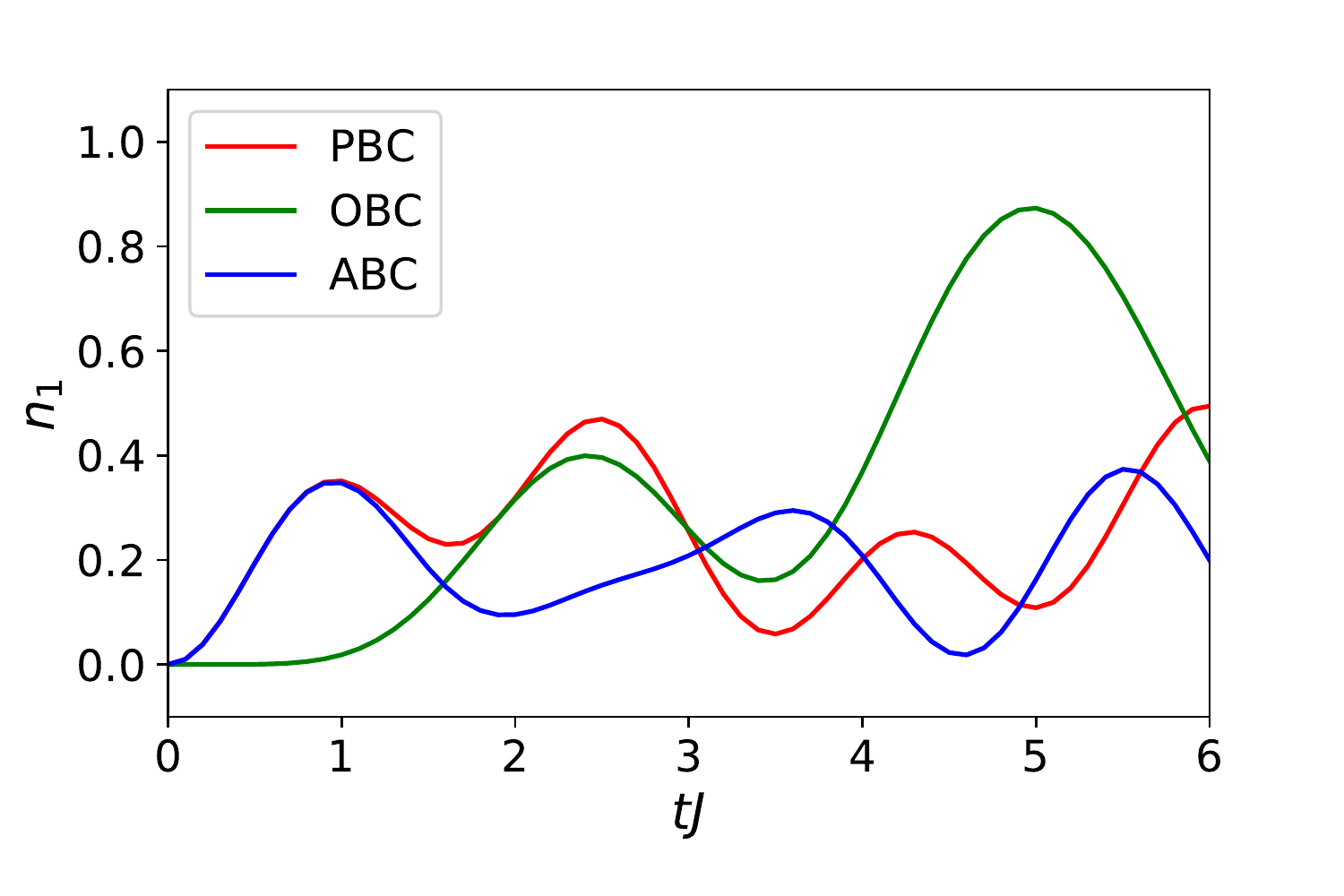}
     \includegraphics[width=0.45\textwidth]{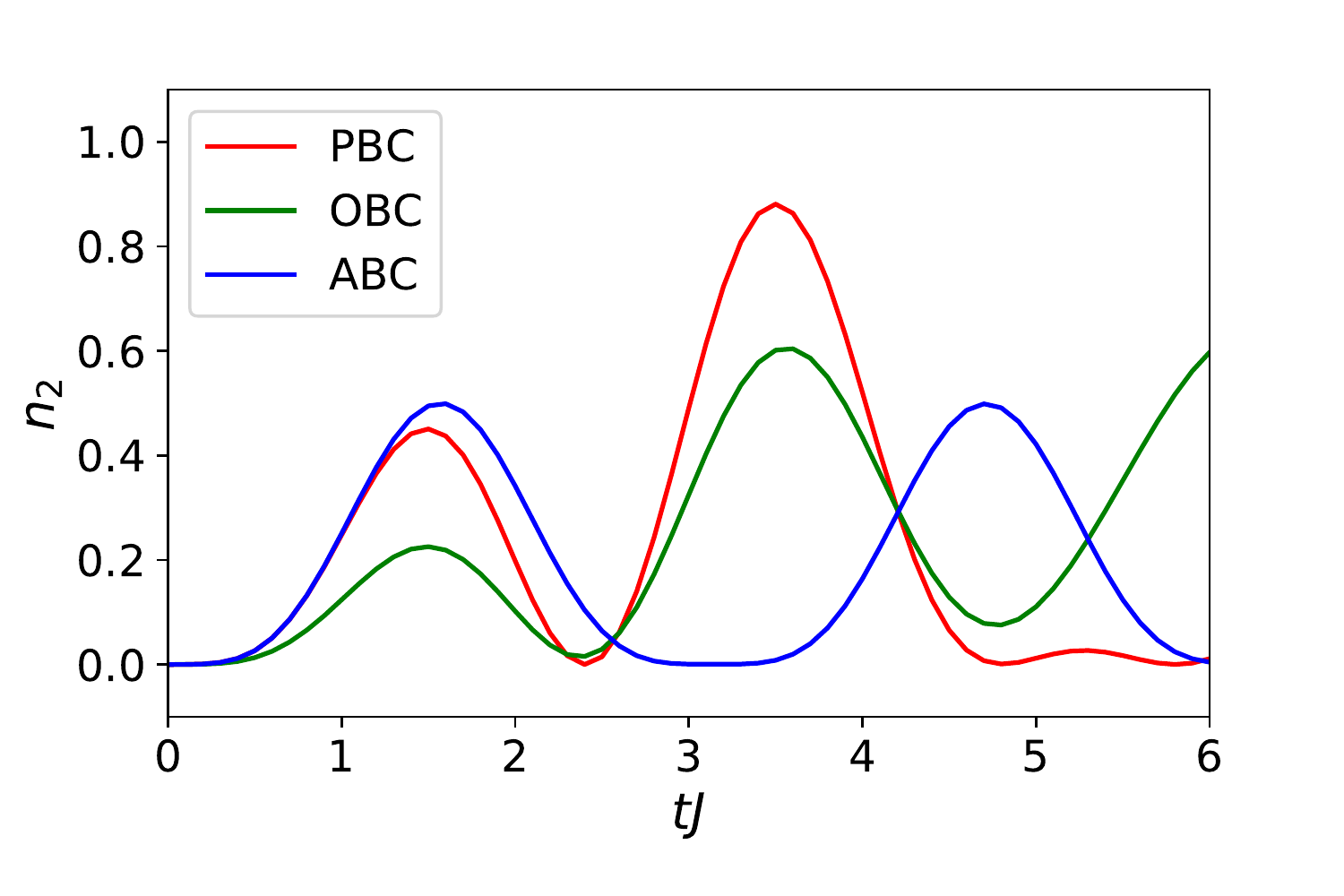} 
     \includegraphics[width=0.45\textwidth]{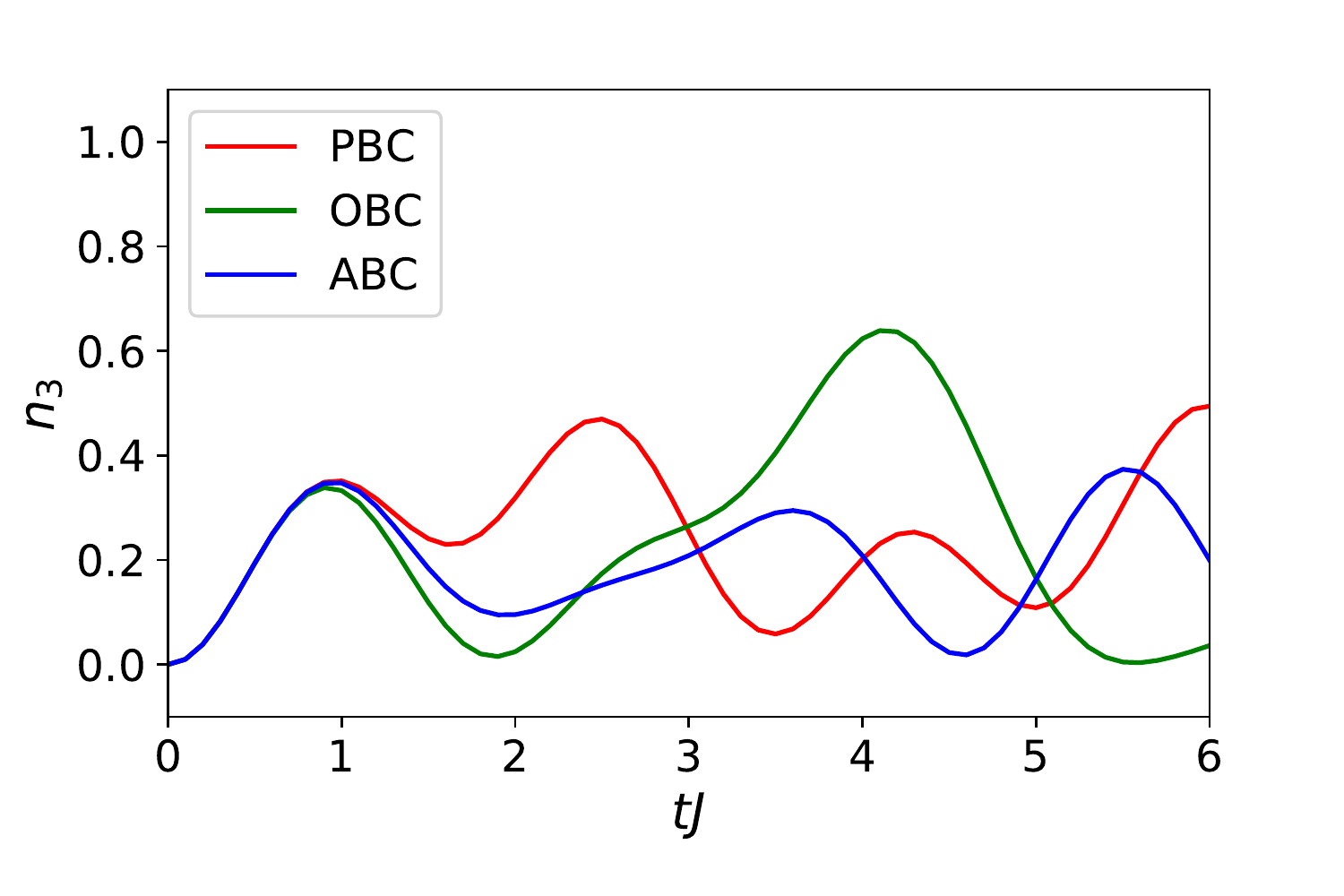} 
     \includegraphics[width=0.45\textwidth]{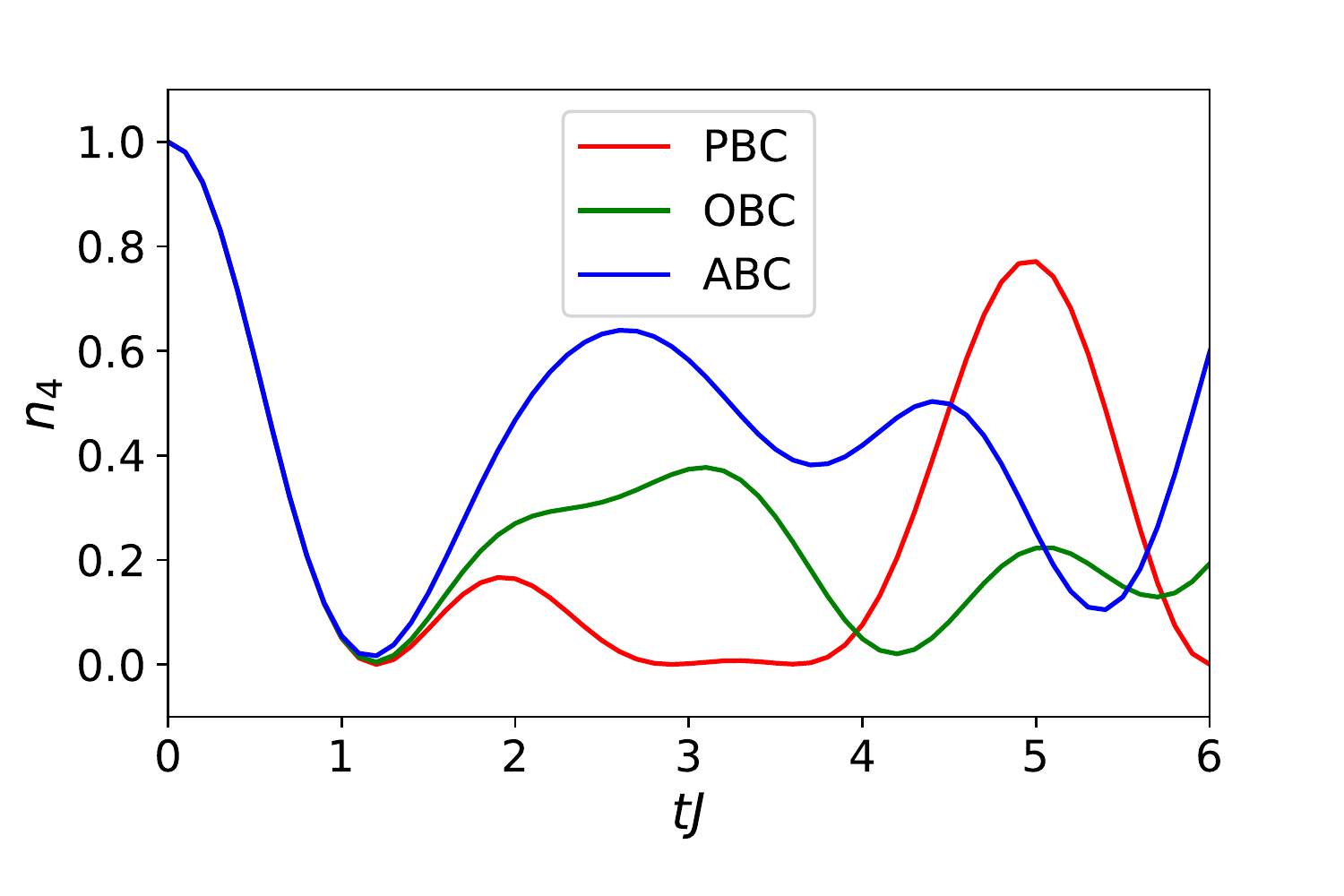}
   \caption{The average particle number, $n_i$, for site $i$, for all four sites as a function of $J t$. Here three different cases of boundary conditions are shown with PBC, OBC and ABC corresponding to periodic, open, and anti-periodic boundary conditions, respectively.}
    \label{fig:compsite}
\end{figure}

The results for one-particle states can be generalized to two-particle states provided that ABC are used. Using lowest order degenerate perturbation theory with ABC we find that the occupation number is, \begin{equation}
    \langle i,j(t)| \hat{n}_l |i,j(t)\rangle \simeq |J^{(N_s)}_{l-i}(2 J t)|^2 + |J^{(N_s)}_{l-j}(2 J t)|^2,
\end{equation}
where $| i  j\rangle = | 0 ... 0, 1_i, 0 ... 0, 1_j,0 ...\rangle$. Agreement is excellent for long time scales, with only a small discrepancy for (c) in Fig. \ref{fig:perturbationcompare}.  It is interesting that after a time, such that $Jt$ is of order 1, the three types of boundary conditions start to give very different values of $\langle n_i(t)\rangle$ (see Fig. \ref{fig:compsite}).
For open boundary conditions, one can compare the situation with that of an ideal gas where the forces exerted on the particles are due to the walls and 
generate the pressure. 

\subsection{Finite volume corrections}
\begin{figure}[H]
    \centering
    \includegraphics[width=0.45\textwidth]{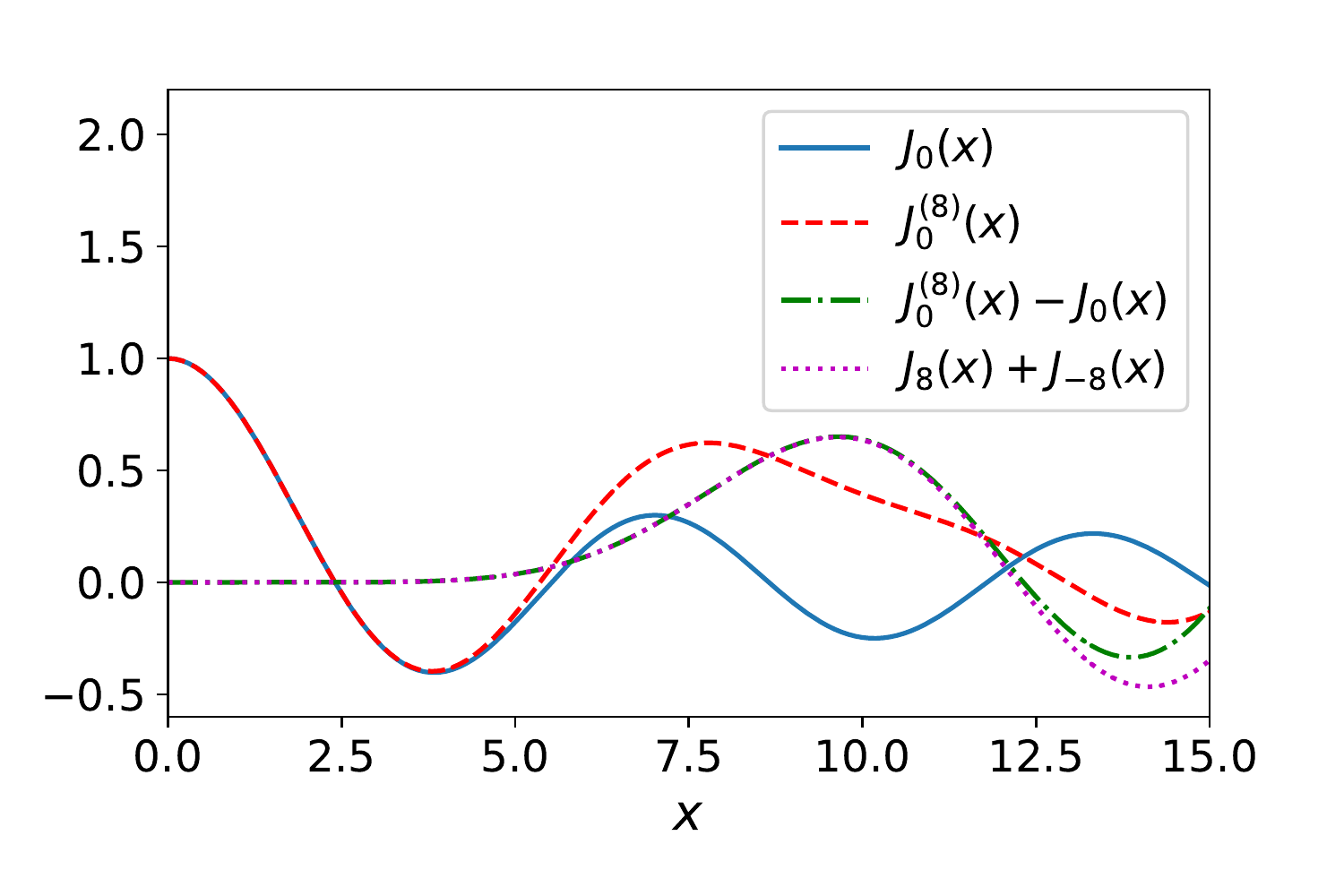}
  \caption{Illustration of the difference between $J_n^{(N_s)}(x)$ and $J_n(x)$  for $N_s=8$ and $n=0$. The solid line represents the usual $J_0(x)$, the dash line the discrete approximation $J_n^{(N_s)}(x)$, the dot-dash line their difference and the dotted line the contribution of the $\ell = \pm 1$ terms in Eq. (\ref{eq:discretebessel}).}
    \label{fig:besselvol}
\end{figure}

It is possible to calculate the difference between the finite volume discrete Bessel functions $J_n^{(N_s)}(x)$ and the usual infinite volume expressions $J_n(x)$. Using the Poisson formula, one finds that
\begin{equation}
\label{eq:discretebessel}
J_n^{(N_s)}(x)=J_n(x)+\sum_{\ell\neq 0}(i)^{N_s\ell}J_{n+N_s\ell}(x),
\end{equation}
where the sum over $\ell$ runs over strictly positive and negative integers.The difference between $J_n^{(N_s)}(x)$ and $J_n(x)$ is small for small argument. This is illustrated for 
$N_s=8$ and $n=0$ in Fig. \ref{fig:besselvol}. One sees that the difference between $J_0^{(8)}(x)$ and $J_0(x)$ 
becomes visible near $x\sim 4$. At that point the difference is almost saturated by the $\ell=\pm1$ terms 
$J_8(x)+J_{-8}(x)$. The $\ell=\pm2$ terms become important near $x\sim 12$.

\section{Formulation of Real Time Evolution}
\label{sec:realtime}

For systems with a small number of spins, the Hilbert space of the model defined by Eq. (\ref{eq:hamiltonianobc}) and Eq. (\ref{eq:hamiltonianpbc}) is small and this evolution operator is tractable for exact implementation on a classical computer; however, the Hilbert space of the model scales like $2^N$ for $N$ spins. For large $N$ ($N \gg 20$) it would only be possible to implement this on a quantum computer as the computational resources instead scale linearly.

Since a quantum computer cannot exactly implement the operator given in Eq. (\ref{eq:evolutionoperator}), we need to use the Suzuki-Trotter (ST) approximation to evaluate the time evolution. We use the first order approximation in order to limit the gate depth of the system:
\begin{equation}
\label{eq:suzukitrotter}
U(\delta t) \simeq e^{i \delta t J \sum \hat{\sigma}^x_{i} \hat{\sigma}^x_{i+1}} e^{i \delta t h_T \sum \hat{\sigma}^z_i} + \mathcal{O}(\delta t^2).
\end{equation}
However we have to apply this operator multiple times to evolve the system to some final time $t$. This iterative process leads to a new expression for the time evolution operator:
\begin{equation}
    \label{eq:suzukitrotterfull}
    U(t;\delta t) \simeq (e^{i \delta t ~ J \sum \hat{\sigma}^x_{i} \hat{\sigma}^x_{i+1}} e^{i \delta t h_T \sum \hat{\sigma}^z_i})^{t/\delta t} + \mathcal{O}((\delta t) ~ t).
\end{equation}

While the estimated worse case error $\mathcal{O}(t ~\delta t )$ is true in general, this bound over estimates the ST truncation error, which should be approximately $\mathcal{O}(J h_T t \delta t)$ because the $\hat{\sigma}^z$ terms only add phase-shifts to the state-vectors of the Hilbert space, and does not affect any measurement of the the basis state. The next order correction to the ST formula is,
\begin{equation}
U^{\text{ST}(2)}(t) =  (e^{i \delta t \frac{h_T}{2} \sum \hat{\sigma}^z_i} e^{i \delta t  J \sum \hat{\sigma}^x_{i} \hat{\sigma}^x_{i+1}} e^{i \delta t \frac{h_T}{2} \sum \hat{\sigma}^z_i})^{t/\delta t}
\end{equation} 
which can be found using the methodology proposed in \cite{Hatano:2005gh}. The second order ST approximation essentially becomes the first order approximation; at this point we can justify the error beginning at the second order ST approximation being $\mathcal{O}(J^3 t( \delta t)^2)$.

\begin{figure}
\centering
	\includegraphics[width=0.5\textwidth]{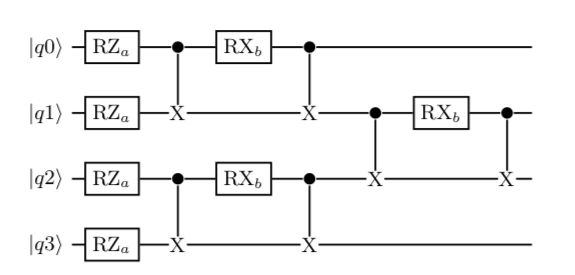}
	\caption{Circuit for 4 qubits with open boundary conditions}
	\label{fig:quantumcircuitobc}
\end{figure}

 It is relatively straightforward to implement the simplest ST approximation as a quantum circuit (see Fig. \ref{fig:quantumcircuitobc}). The Hamiltonian is split as follows: $H = H_1 + H_2  + H_3$, where,
 \begin{equation}
 \begin{split}
 H_1 &= -h_T \sum_{i = 1}^N \sigma^z_i \\
 H_2 &= -J \sum_{i = 1,~3,~5...}^N \sigma^x_i \sigma^x_{i+1}\\
 H_3 &= -J \sum_{i = 2,~4,~6...}^N \sigma^x_i \sigma^x_{i+1}.\\
 \end{split}
 \end{equation} 
 The $H_1$ term can easily be implemented as a single moment in a quantum circuit. While $H_2$ and $H_3$ commute, they have to be implemented as separate moments in the quantum circuit because each of these contain terms which entangle two qubits. This corresponds to the idea \cite{Lloyd1073} of splitting the Hamiltonian into pieces that can be implemented easily separately and when combined correspond to the original Hamiltonian with a Trotter error.  On the other hand, the on-site terms can be executed in a single moment in a quantum circuit as these are single qubit operators.  
The implementation of this circuit for an arbitrary number of qubits has a gate depth ($d_l$),
\begin{equation}
\label{eq:depth}
    d_l = 7 * N_t,
\end{equation}
and a total number of gate operations:

\begin{equation}
\label{eq:ngatespbc}
N_{\text{gates}}^{\text{pbc}} = 2* (N_s*2)*N_t
\end{equation}
\begin{equation}
\label{eq:ngatesobc}
\begin{split}
N_{\text{gates}}^{\text{obc}} &= (N_s)*2 - 1)*N_t\\
&+ ((N_s - 1)*2)*N_t.
\end{split}
\end{equation}
where $N_s$ is the number of sites and $N_t$ is the number of trotter steps. 

\section{Results of Real Time Evolution}
\label{sec:results}

We examined two different cases of the 1-D Ising model with 8 sites (with both OBC and PBC): the time evolution of a single particle initial state and scattering of two particles. For all cases we examined the system with the nearest neighbor coupling $J = 0.02$ and on-site coupling $h_T = 1.0$. We define the initial states for the system in Table \ref{tab:initstates}. 
\begin{table}[H]
    \centering
    \begin{tabular}{|c|c|c|}
    \hline
        Sim Type & OBC & PBC \\ \hline
        Free Prop & $\ket{10000000}$ & $\ket{10000000}$\\ \hline
        Scatt & $\ket{10000001}$ & $\ket{10001000}$\\ \hline
    \end{tabular}
    \caption{Initial State of the system}
    \label{tab:initstates}
\end{table}

\subsection{Sources of error}
\begin{figure*}

	\renewcommand{\thesubfigure}{\Alph{subfigure}}
	\begin{subfigure}[H]{0.49\textwidth}
	
    		\includegraphics[width=\textwidth]{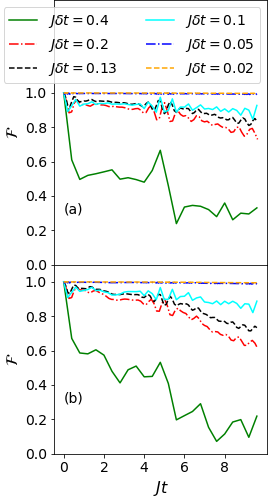}
    		\caption{open boundary conditions}
    		\label{fig:obcfidelity}

	\end{subfigure}
	\begin{subfigure}[H]{0.49\textwidth}
    		\includegraphics[width=\textwidth]{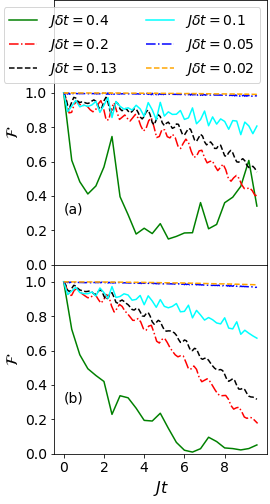}
    		\caption{periodic boundary conditions}
    		\label{fig:pbcfidelity}

	\end{subfigure}
	\caption{Fidelity of the Trotter operator at multiple different Trotter steps for (A) free propagation and (B) scattering with different boundary conditions and $J = 0.02$, $h_T = 1.0$ and $N_s = 8$}
\end{figure*}

Because the ST approximation is an iterative process we want to know how many operations can be carried out before imperfections in the approximations and the noisy gates of a quantum computer will produce substantial issues with our simulations. A first step is to measure the fidelity of the Trotter operator with the exact evolution operator over the time scales of interest in these processes (see Fig. \ref{fig:obcfidelity} and Fig. \ref{fig:pbcfidelity}). The fidelity of the ST operator is,
\begin{equation}
    \label{eq:fidelity}
    \mathcal{F}(t;\delta t) = |\bra{\psi(0)} (U^{\text{exact}}(t))^{\dagger} U^{ST}(t;\delta t) \ket{\psi(0)}|.
\end{equation}
%

For both OBC and PBC it appears that a trotter step of $\delta t = 10.0$ for $J = 0.02$ is satisfactory to describe the time evolution for small time scales.

\begin{figure}
    \centering
    \includegraphics[width=0.4\textwidth]{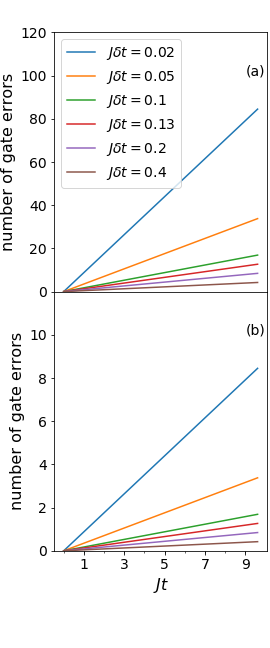}
    \caption{Average number of gate errors ($y$-axis) for state of the art Trapped Ion (a) and expected near term noisy digital quantum computer (b)}
    \label{fig:gateerrors}
\end{figure}

It is important to have an understanding of how frequent quantum gate errors will be when applying the ST operator.  We expect the number of gate errors to increase the more Trotter steps we apply. In Fig. \ref{fig:gateerrors}, we show the average number of gate errors (Pauli Channel) at a given evolution time for a state of the art trapped ion system ($p_{\text{1 qubit}} = 1.0*10^{-4}$ and $p_{\text{2 qubit}} = 5.0*10^{-4}$~\cite{PhysRevLett.113.220501,PhysRevLett.117.060504}) and for slight improvements of current typical digital quantum computers~\cite{spec:IBM,spec:Rigetti}

The Pauli error channel for single-qubit gates is defined in terms of the density matrix $\hat{\rho}$:
\begin{equation}
    \label{eq:paulichannel}
    \mathcal{E}(\hat{\rho};p_x,p_y,p_z) = (1 - p) \hat{\rho} + p_x \hat{\sigma}^x \hat{\rho} \hat{\sigma}^x + p_y \hat{\sigma}^y \hat{\rho} \hat{\sigma}^y + p_z \hat{\sigma}^z \hat{\rho} \hat{\sigma}^z.
\end{equation}
The values $p_x$, $p_y$, and $p_z$ correspond to the probabilities of an $\sigma^x$, $\sigma^y$, or $\sigma^z$ error respectively occurring and $p = p_x + p_y + p_z$. The error channel for two qubit gates is given by $\mathcal{E}^{(2)} = \mathcal{E}\bigotimes \mathcal{E}$. The values of $p_x,~p_y,~p_z$ for the one and two-qubit gates are given in Tables \ref{tab:errorrates} and \ref{tab:errorrates2}. We also introduce measurement errors into our simulations. These are caused by misidentifying the state that the qubit is in (i.e. reading a 1 as a 0 or a 0 as a 1). We implemented this by changing the readout value with a chance $p_{\text{measure}}$ given in Tables \ref{tab:errorrates} and \ref{tab:errorrates2}. Ref. \cite{Kandala:2017aa} identifies a procedure to address the readout error in the supplementary material; we simplify their result to using the following rescaling of the measured readout:
\begin{equation}
\label{eq:readoutfix}
\langle Z^{\text{exact}} \rangle = \frac{\langle Z^{\text{noisy}} \rangle}{p_{\text{measure}}}
\end{equation}
\begin{table}[H]
    \centering
    \begin{tabular}{|c|c|c|}
    \hline
        pauli ch. & 1 qubit error & 2 qubit error \\
        \hline
        $p_x,~p_y$ & 0.00002 & 0.0001\\ \hline
        $p_z$ & 0.00006 & 0.0003 \\ \hline
        meas. ch. & 1 qubit error & 2 qubit error \\ \hline
        $p_{\text{measure}}$ & 0.05 & - \\ \hline
        
    \end{tabular}
    \caption{Optimistic error rates for current trapped ions}
    \label{tab:errorrates}
\end{table}

\begin{table}[H]
    \centering
    \begin{tabular}{|c|c|c|}
    \hline
        pauli ch. & 1 qubit error & 2 qubit error \\
        \hline
        $p_x,~p_y$ & 0.00033 & 0.0033\\ \hline
        $p_z$ & 0.00033 & 0.0033 \\ \hline
        meas. ch. & 1 qubit error & 2 qubit error \\ \hline
        $p_{\text{measure}}$ & 0.05 & - \\ \hline
        
    \end{tabular}
    \caption{Optimistic error rates for near term superconducting qubits}
    \label{tab:errorrates2}
\end{table}

\subsection{Methods for dealing with error}
Approximating the time evolution operator using the ST method introduces an error $\mathcal{O}(\delta t ~ t)$. It was suggested in \cite{2018arXiv180803623E,PhysRevX.8.031027,PhysRevX.7.021050} that by using simulations at multiple different Trotter steps and noise levels it is possible to systematically reduce the uncertainty in the measured quantities. 

In order to minimize the noise error, Ref. \cite{PhysRevX.8.031027} suggested several methods, one of which is an exponential extrapolation,
\begin{equation}
    \label{eq:expnoise}
    \langle O \rangle (0) = (\langle O(\epsilon)\rangle)^{\frac{r}{r-1}} (\langle O(\epsilon)\rangle)^{\frac{1}{1-r}}.
\end{equation} In Eq. (\ref{eq:expnoise}), $\epsilon$ is the noise rate for the system which is dependent upon the probability of gate errors occurring as in Eq. (\ref{eq:paulichannel}) and $r$ is a scale factor such that $r>1$.  Due to the computational overhead of carrying out this methodology of error mitigation we only demonstrate a modification of it in our results for free propagation on a four site lattice. Our modification to the error mitigation scheme proposed in \cite{2018arXiv180803623E,PhysRevX.8.031027,PhysRevX.7.021050} involves changing the extrapolation equation proposed in Eq. \eqref{eq:expnoise} to an exponential ansatz of the form:
\begin{equation}
\label{eq:exponentialansatz}
\langle \mathcal{O}(\epsilon*r;t) \rangle = A B^r + C.
\end{equation} 
This form should retain the same general behaviors of the ansatz  proposed in Eq. \eqref{eq:expnoise}. 

The methods for reducing algorithmic errors are more computationally intensive and discussed in Ref. \cite{2018arXiv180803623E}. They argue that the algorithmic error rate, $\epsilon_N$, scales as $1/N$, where N is the number of trotter steps. Due to the increased computational demand of this error mitigation method we currently have not implemented it. 

\subsection{Simulation Results }
\begin{figure*}
	\renewcommand{\thesubfigure}{\Alph{subfigure}}
	\begin{subfigure}[H]{0.45\textwidth}
   		 \includegraphics[width=\textwidth]{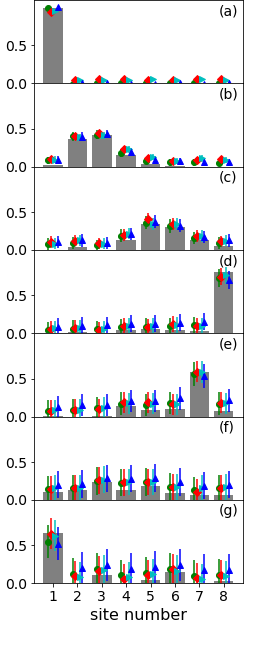}
		 \caption{open boundary conditions}
   		 \label{fig:obcfreepropogation}
    	\end{subfigure}
	\begin{subfigure}[H]{0.45\textwidth}
   		 \includegraphics[width=\textwidth]{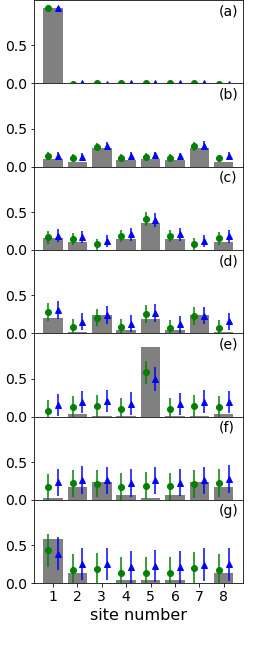}
   		 \caption{periodic boundary conditions}
   		 \label{fig:pbcfreeprop}
	\end{subfigure}
	\caption{Free propagation with $J=0.02$, $h_{T}=1.0$, $J\delta t = 0.4$ with 10000 shots at time steps: (a) $Jt = 0$, (b) $Jt = 1.6$, (c) $J t=3.2$, (d) $Jt = 4.8$, (e) $Jt = 6.4$, (f) $Jt = 8.0$, (g) $Jt = 9.6$. Green circles: QISKIT simulation for current trapped ions, red diamonds: numpy simulation for current trapped ion, blue triangle: QISKIT simulation for near future superconducting qubit quantum computers, cyan right arrow: numpy simulation for near future superconducting qubit quantum computers, gray bars: exact diagonalization}
\end{figure*}

\begin{figure*}
	\renewcommand{\thesubfigure}{\Alph{subfigure}}
	\begin{subfigure}[H]{0.45\textwidth}
   		 \includegraphics[width=\textwidth]{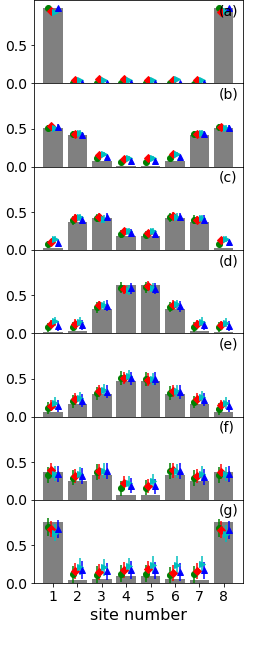}
		\caption{open boundary conditions}
   		 \label{fig:obcscattering}
	\end{subfigure}
	\begin{subfigure}[H]{0.45\textwidth}
    		\includegraphics[width=\textwidth]{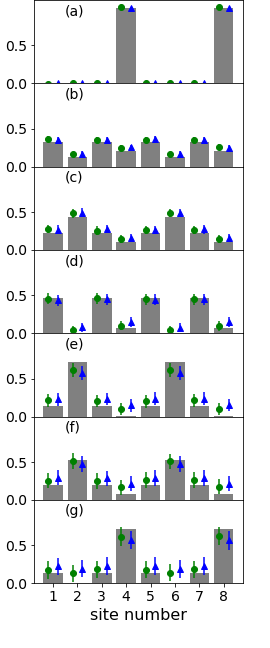}
    		\caption{periodic boundary conditions}
   		 \label{fig:pbcscattering}
	\end{subfigure}
	\caption{Scattering with $J=0.02$, $h_{T}=1.0$, $J\delta t = 0.4$ with 10000 shots at time steps: (a) $Jt = 0$, (b) $Jt = 0.8$, (c) $J t=1.6$, (d) $Jt = 2.4$, (e) $Jt = 3.2$, (f) $Jt = 4.0$, (g) $Jt = 4.8$. Green circles: QISKIT simulation for current trapped ions, red diamonds: numpy simulation for current trapped ion, blue triangle: QISKIT simulation for near future superconducting qubit quantum computers, cyan right arrow: numpy simulation for near future superconducting qubit quantum computers, gray bars: exact diagonalization}
\end{figure*}

The initial state of the system for each case we looked at is given in Table \ref{tab:initstates}. We evolved the system using exact diagonalization of the Hamiltonian as well as simulated using a quantum virtual machine implemented through the python library QISKIT published by IBM, which implemented the noise corresponding to the Pauli channel for the system. The results from the QISKIT simulations were checked for consistency by implementing the noise channel using matrix operations instead of the QISKIT library. The results of the simulation are show in Fig. \ref{fig:obcfreepropogation} through Fig. \ref{fig:pbcscattering}. We use a reduced $\chi^2$ value ($\tilde{\chi}^2$) as a metric for how well the simulated quantum computer predicts the exact results, we do this by comparing the difference at different sites,
\begin{equation}
\tilde{\chi}^2(t) =\frac{1}{N} \sum_{j = 1}^{N} \Big(\frac{\langle n_j^{\text{exact}}(t) \rangle - \langle n_j^{\text{sim.}}(t) \rangle}{\delta \langle n_j^{\text{sim}} \rangle}\Big)^2.
\end{equation}
These values over time are given in Table \ref{tab:chi2timeprop} and Table \ref{tab:chi2timescat}. The systematic errors from the ST approximation become significant at time scales on the order of $Jt \approx 6$. These results suggest that it is possible to simulate a real time scattering event and measure the matrix elements of a correlation function, for a simple field theory. 

\begin{table}[H]
\centering
\begin{tabular}{| c | c | c | }
\hline
metric & obc & pbc \\ \hline
$\tilde{\chi}^2(0)$ & 1.5 & 0.91\\ \hline
$\tilde{\chi}^2(80)$ & 4.41 & 0.5\\ \hline
$\tilde{\chi}^2(160)$ & 2.43 & 0.45\\ \hline
$\tilde{\chi}^2(240)$ & 4.4& 0.2\\ \hline
$\tilde{\chi}^2(320)$ & 0.88 & 0.13 \\ \hline
$\tilde{\chi}^2(400)$ &3.2 & 0.28\\ \hline
$\tilde{\chi}^2(480)$ & 4.1& 0.33\\ \hline
\end{tabular}
\caption{$\tilde{\chi}^2$ at different times over all sites for free propagation}
\label{tab:chi2timeprop}
\end{table}

\begin{table}[H]
\centering
\begin{tabular}{| c | c | c |}
\hline
metric & obc scattering & pbc scattering  \\ \hline
$\tilde{\chi}^2(0.0)$&2.1& 	 1.7  \\ \hline
$\tilde{\chi}^2(40.0)$&7.4 & 	0.71      \\ \hline
$\tilde{\chi}^2(80.0)$&4.3 & 	 0.525        \\ \hline
$\tilde{\chi}^2(120.0)$&1.4& 	  0.19     \\ \hline
$\tilde{\chi}^2(160.0)$&2.9& 	  1.1    \\ \hline
$\tilde{\chi}^2(200.0)$&9.1& 	 0.38    \\ \hline
$\tilde{\chi}^2(240.0)$&12.9& 	0.6   \\ \hline
\end{tabular}
\caption{$\tilde{\chi}^2(t)$ for scattering and mixed field simulations for 8 sites; the number of degrees of freedom is 8}
\label{tab:chi2timescat}
\end{table}

\subsection{Results of simulated superconducting qubit quantum computer}
\begin{figure}
\includegraphics[width=0.49\textwidth]{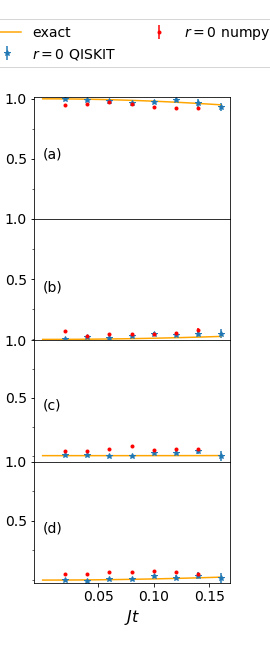}
\caption{Plot of occupation of different sites as a function of $Jt$ for 4 site propogation simulation with $J = 0.02$, $h_T = 1.0$, $J~\delta t = 0.02$, (a) site 1, (b) site 2, (c) site 3, (d) site 4.}
\label{fig:dt1propogation}
\end{figure}

\begin{figure}
\includegraphics[width=0.49\textwidth]{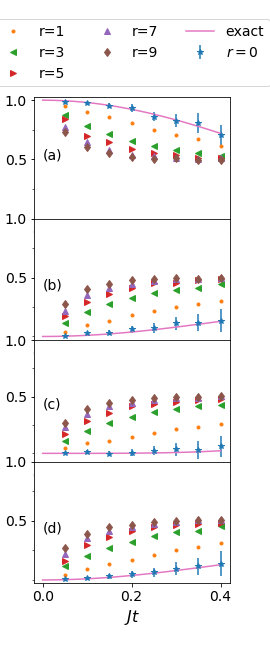}
\caption{Plot of occupation of different sites as a function of $Jt$ for 4 site propogation simulation with $J = 0.02$, $h_T = 1.0$, $J~\delta t = 0.05$, (a) site 1, (b) site 2, (c) site 3, (d) site 4.}
\label{fig:dt25propogation}
\end{figure}

\begin{figure}
\includegraphics[width=0.49\textwidth]{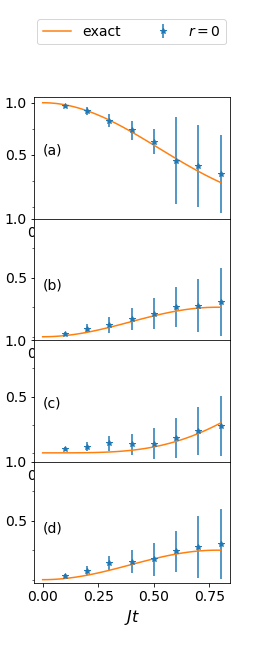}
\caption{Plot of occupation of different sites as a function of $Jt$ for 4 site propogation simulation with $J = 0.02$, $h_T = 1.0$, $J~\delta t = 0.1$, (a) site 1, (b) site 2, (c) site 3, (d) site 4.}
\label{fig:dt50propogation}
\end{figure}

In addition it would be interesting to see if it is possible to extract results using current superconducting qubit quantum computers. In this case we used $p_x=p_y=p_z=0.0005$ for the one qubit gates and $p_x=p_y=p_z=0.004$ for the two qubit gates; this corresponds to a gate error of $\sim$ $0.01$ for one qubit gates and $\sim 0.04$ for two qubit gates. In this case we need to address the issues of noisy quantum gates because of the number of two qubit gates we have in our system.  To do this we simulated the system at 4 different noise levels by introducing noisy identity operators into our circuit. Similarities can be seen between Fig. \ref{fig:dt25propogation} and Fig. 4 in Ref. \cite{Klco:2018kyo}. The only difference in the procedure that we carried out is the extrapolation method that we used. Ref.~\cite{Klco:2018kyo} used a quadratic ansatz while we used an exponential ansatz of the form:
\begin{equation}
\langle \mathcal{O}(\epsilon*r;t) \rangle = A B^r + C
\end{equation}
to extrapolate the noiseless observable. We used priors of $A = 0.0\pm0.5$, $B=0.0\pm1.0$, and $C=0.5\pm0.5$. For proof of concept we worked at 4 sites with $J=0.02$ and $h_T=1.0$ and we took 8000 measurements for each data point at each noise level. The results of these simulations are shown in Figs. \ref{fig:dt1propogation} and \ref{fig:dt25propogation}. The errors found at later times in Fig. \ref{fig:dt50propogation} are likely due to the signal and the noise level being so close together that it is difficult for a fit to yield an accurate noiseless extrapolation. 

\subsection{Continuation to larger J}
We have examined tentatively regions where $J = 0.2$ and $h_T = 1.0$. The most noticeable effect is the particles hop between sites far more quickly. While this case is still far away from the continuum limit, i.e. $J = h_T$, in this regime pair creation and annihilation becomes more frequent. This implies that the particle conservation picture breaks down quicker and second order effects in degenerate perturbation theory become significant. In addition we find that the particle occupation at given sites is not as regular as in the case where $J << h_T$. 

The most significant change in working with a larger value of $J$ is the Trotter time steps must be shrunk because the Trotter truncation error still scales in a similar manner as the regime $J=0.02$. We expect some issues arising from noisy simulations will be similar to those encountered in the $J\delta t = 0.1$ shown in Fig. \ref{fig:dt50propogation}, with noiseless extrapolation. Specifically when the lowest noise simulation observable is close to, or crosses, the observed value for inflated noise simulations which no longer have a discernible signal, the noiseless extrapolation method produces substantially larger uncertainties for the observable.

\section{conclusion}
We have demonstrated through simulations on an emulated quantum computer that it is possible to use current trapped ion systems to simulate the real-time evolution of the quantum Ising model with both 4 and 8 sites, and in the near future it will be possible to be simulated on quantum computers using superconducting qubits. Currently, the density matrix renormalization group and tensor networks are the only methods we have of examining real-time scattering; however, in the near future quantum computers will be able to join this group of tools so that we can examine these systems in real time. We have derived a simple perturbative expression that can be used to check the consistency of the results done on a system of trapped ions or in the near future on superconducting qubits. These perturbative expressions can be used for much larger systems and can be easily handled analytically and numerically.

Much work remains to be done in order to study the real-time evolution of interacting particles close to the continuum limit. We plan to examine related theories, such as the $O(3)$ non-linear sigma model where the triplet and singlet states could be implemented with a pair of qubits, or slight modifications to the Ising model such as changes in the transverse field, because these models allow us to examine a richer volume of observables such as phase shifts, scattering cross sections, and bound states. 

\begin{acknowledgments}
This work was supported in part by the U.S. Department of Energy (DOE) under Award Number DE-SC0019139. We thank the members of this grant for stimulating discussions. In particular we thanks Stephen Jordan, Nathalie Kclo, and Martin Savage for their input. We thank Jacques Perk for the suggestion of references regarding the XY-model. JUY was supported by the US Department of Energy (DOE), Office of Science, Office of High Energy Physics, under Award Numbers DE-SC0009998.
\end{acknowledgments}

\newpage
\input{mainrevtexv2.bbl}
\end{document}

%% file: mainrevtexv2.bbl
%